\documentclass[11pt]{article}
\usepackage{amsmath, mathrsfs, amsthm, amsfonts}
\usepackage{amssymb}

\usepackage{booktabs}
\usepackage{mathtools}
\usepackage{hyperref}
\usepackage{soul}
\usepackage{dsfont}
\usepackage{bm}
\usepackage[round]{natbib}
\usepackage[capitalise,noabbrev]{cleveref}
\usepackage[plain,noend]{algorithm2e}
\usepackage{subcaption}
\usepackage{float}
\usepackage{graphicx}
\usepackage{enumitem}
\usepackage{bbold}
\usepackage{xcolor}
\usepackage{color}
\graphicspath{{../../Images/normalizing-flow/}}

\newcommand{\indep}{\perp \!\!\! \perp}

\newcommand{\be}{\begin{equation*}\begin{aligned} }
\newcommand{\ee}{\end{aligned}\end{equation*} }

\newcommand{\bel}{\begin{equation}\begin{aligned} }
\newcommand{\eel}{\end{aligned}\end{equation} }
\renewcommand{\t}[1]{\text{#1} }

\begin{document}

\title{Normalizing Flow to Augmented Posterior:\\ Conditional Density Estimation with Interpretable Dimension Reduction for High Dimensional Data}

\author{Cheng Zeng
	\thanks{Department of Statistics, University of Florida, U.S.A. czeng1@ufl.edu}\quad\quad
	George Michailidis
	\thanks{Department of Statistics and Data Science, University of California Los Angeles, U.S.A. gmichail@ucla.edu}\quad\quad
	Hitoshi Iyatomi
	\thanks{Department of Applied Informatics, Graduate School of Science and Engineering, Hosei University, Japan. iyatomi@hosei.ac.jp}\quad\quad
	Leo L Duan
	\thanks{Corresponding author. Department of Statistics, University of Florida, U.S.A. \href{email:li.duan@ufl.edu}{li.duan@ufl.edu}}
}

\maketitle

\begin{abstract}
	The conditional density characterizes the distribution of a response variable $y$ given other predictor $x$, and plays a key role in many statistical tasks, including classification and outlier detection. Although there has been abundant work on the problem of Conditional Density Estimation (CDE) for a low-dimensional response in the presence  of a high-dimensional predictor, little work has been done for a high-dimensional response such as images.
	The promising performance of normalizing flow (NF) neural networks in unconditional density estimation acts a motivating starting point. In this work, we extend NF neural networks when external $x$ is present. Specifically, they use the NF to parameterize a one-to-one transform between a high-dimensional $y$ and a latent $z$ that comprises two components \([z_P,z_N]\).
	The $z_P$ component is a low-dimensional subvector obtained from the posterior distribution of an elementary predictive model for $x$, such as logistic/linear regression. The $z_N$ component is a high-dimensional independent Gaussian vector, which explains the variations in $y$ not or less related to $x$. Unlike existing CDE methods, the proposed approach, coined Augmented Posterior CDE (AP-CDE), only requires a simple modification on the common normalizing flow framework, while significantly improving the interpretation of the latent component, since $z_P$ represents a supervised dimension reduction. In image analytics applications, AP-CDE shows good separation of $x$-related variations due to factors such as lighting condition and subject id, from the other random variations. Further, the experiments show that an unconditional NF neural network, based on an unsupervised model of $z$, such as Gaussian mixture, fails to generate interpretable results.
\end{abstract}


\section{Introduction}

A conditional density characterizes the probabilistic behavior of a set of random variables, when information on a set of other variables is available. The case of a single variable $y$ (the response) conditioned on a multivariate $x$ (predictor) has received most attention in the literature, due to a wide range of applications. A number of methods have been proposed to address the conditional density estimation (CDE) problem from observed data. Kernel density \citep{terrell1992variable, botev2010kernel,kim2012robust} and k-nearest neighbors  \citep{mack1979multivariate,kung2012optimal} based techniques have been extensively studied and employed in applications. Another popular approach uses a mixture model of the form $f(y_i \mid x_i) = \sum_{k=1}^K w_k(x_i) g(y_i \mid \theta_k(x_i))$ over data index $i=1,\ldots,n$, with $\sum_{k=1}^K w_k(x_i)=1$ and potentially $K\to\infty$. In the mixture model and for continuous $y_i$, $g(\cdot)$ can be a location-scale density, such as a multivariate Gaussian one with mean $\mu_k$ and covariance matrix $\Sigma_k$. Importantly, the mixture component parameters $\theta_k$ as well as the mixture weights $w_k$ are some deterministic transforms of the $x_i$'s, hence allowing the mixture distribution of $y_i$ to vary according to $x_i$. There is a large literature in such a framework, see, e.g.,
\cite{jiang1999hierarchical,geweke2007smoothly,villani2009regression,  norets2010approximation,huang2022evaluating} and references therein.

The CDE framework has a wide range of statistical applications, besides serving as a nonlinear predictive model for $y_i$, including outlier detection and classification. In the first case, by evaluating the magnitude of $f(y_i\mid x_i)$ for each observed data point, one could identify those points in the bottom density quantile as potential outliers \citep{scott2004partial,schubert2014generalized}. In the latter case, when $x_i$ is a discrete class label, such as whether a patient is in disease status, with a class probability $p(x_i)$, $p(x_i\mid y_i)\propto p(x_i)f(y_i\mid x_i)$ could be used as a probabilistic classifier for predicting $x_i$ \citep{garg2001understanding}. 

Despite significant advances in CDE, in recent years, a number of major challenges are not satisfactorily addressed for a high-dimensional response $y_i\in\mathbb{R}^p$, such as an image. Hence, the focus of the current work is on the case of a high-dimensional response $y_i$---rather than low-dimensional response conditioned on a high-dimensional predictor $x_i$  in extant literature. 
For the latter, a large class of solutions exist, such as BART \citep{chipman2010bart}, that partition the high-dimensional space of $x_i$ and use simple piece-wise distribution (such as spherical Gaussian, or Bernoulli) for the low-dimensional $y_i$ in each region. Obviously, this strategy does not be applied to high-dimensional $y_i$.

When using the mixture models for density estimation, the following two difficulties arise for a high-dimensional data $y$: (i) specification of the component distribution $g$ in the mixture model and (ii) the curse of dimensionality when computing a high-dimensional mixture distribution. These two points are elaborated next. First, a parametric specification of the component distribution $g$ can be unsatisfactory, since location-scale distribution is often an over-simplification for  high-dimensional data. For example, for a collection of face photos from one subject, the mean of a Gaussian density $g$ is often a poor summary characterization for this group of data points, since there are other factors (such as unknown lighting conditions) that contribute to the within-group variability, besides just pixel-wise random noise. To address this issue, it is often useful to assume that the high-dimensional data point lie close to several manifolds, each having an intrinsic low dimension. One of the well-known solutions is the mixture of factor analyzers (MFA) \citep{mclachlan2000mixtures, tang2012deep}, or mixture of probabilistic principal component analysis (PCA) \citep{tipping1999mixtures}, which retains the multivariate Gaussian density $g(y_i\mid \mu_k(x_i), \Sigma_k)$; further, the covariance matrix takes the form $\Sigma_k=U_k U'_k + \Lambda_k$, with $U_k$ some $p\times d$ matrix with $d$ small, and $\Lambda_k$ a diagonal positive matrix. Effectively this parameterization assumes that for some $h$, $y_i-\mu_k(x_i)$ lies near a linear subspace spanned by the columns of $U_k$. However, it is rather difficult to extend this technique to non-linear manifolds through non-linear mappings.
Alternatively, a popular idea is to find a low-dimensional representation, or ``embeddings'' that preserve some (often not all) relational characteristics of the high-dimensional data, such as pairwise distances or local neighborhoods. Such embedding is denoted by $z_i\in \mathbb{R}^d$ for each data point.
Examples include Sammon's mapping \citep{sammon1969nonlinear}, kernel PCA \citep{scholkopf1997kernel}, Laplacian eigenmaps \citep{belkin2003laplacian}, locally-linear embeddings \citep{roweis2000nonlinear}, Gaussian process latent variables (GPLV) \citep{lawrence2003gaussian}, t-distributed stochastic neighbor embeddings (t-SNE) \citep{van2008visualizing}, uniform manifold approximation and projection (UMAP) \citep{mcinnes2018umap}, just to name a few. After obtaining such $z_i$'s, one could calculate the  conditional density $f(z_i\mid x_i)$ for $z_i$, as a surrogate the corresponding conditional density for $y_i$. Despite some success in visualization tasks and cluster analysis, a key issue among the aforementioned methods is that the procedure of dimension reduction often lacks a generative distribution (except for GPLV); consequently, a density for $f(y_i\mid x_i)$ can not be obtained.

The second difficulty is the curse of dimensionality when computing a high-dimensional mixture distribution, which is often overlooked in the literature. Common algorithms used for mixture model estimation involve a discrete latent variable $z_i=h$ with probability $w_k$, given that $(y_i \mid z_i=h, x_i)$ comes from a component distribution $g(\cdot \mid \theta_k(x_i))$.  Since $p(z_i=h \mid y_i, \theta_k)\propto w_k f(y_i \mid \theta_k(x_i))$, it allows iterating through the following two steps: (a) sampling of $z_i$ via a multinomial distribution (or taking expectation $\tilde z_i= p(z_i=h \mid y_i, \theta_k)$ in the EM algorithm \citep{fraley2002model}); (ii) updating $\theta_k(x_i)$ conditioned on $z_i$ or $\tilde z_i$. Nevertheless, since for $h=1,\ldots, K$, the high-dimensional $y_i$ creates large magnitude of densities, thus resulting in $f(y_i\mid \theta_k(x_i))/f(y_i\mid \theta_{l}(x_i))\approx 0$ or $\infty$, unless $\theta_{k}(x_i)\approx \theta_l(x_i)$. Consequently, each $p(z_i\mid y_i,\theta_k(x_i))$ is very likely to be stuck at 1 for a given $h^*$, and close to 0 for all the other $h\neq h^*$, with the end result being that the estimation algorithm ``gets stuck" at the initial assignment $z_i$'s.

The curse-of-dimensionality for density estimation is a well known issue and the reason that algorithms such as rejection sampling, importance sampling, and Metropolis-Hastings fail in high dimensions \citep{gelfand2000gibbs,zuev2011optimal}. A similar problem was recently discovered in high-dimensional clustering \citep{chandra2023escaping}, which is closely related to the unconditional density estimation problem.

These challenges in high-dimensional density estimation approaches motivated the development of completely different approaches. Normalizing flow neural networks were proposed to find an invertible mapping between a random variable $y_i\in \mathbb{R}^p$ and a latent variable $z_i \sim \text{N}(0,I_p)$. The neural network is formed by stacking layers of non-linear transforms, each layer parameterized in the way such that it corresponds to a bijective transform, and the inverse transform has a closed-form or can be computed efficiently.
Using a simple change-of-variable technique, one could obtain the density $f(y_i)$ as a transformed density from an independent Gaussian one. Examples include RealNVP \citep{dinh2016density}, MADE \citep{germain2015made}, MAF \citep{papamakarios2017masked}, Glow \citep{kingma2018glow}, FFJORD \citep{grathwohl2018ffjord} and iResNet \citep{behrmann2019invertible,chen2019residual}. Due to the large number of parameters and expressiveness of neural networks, impressive performance has been exhibited  as a generative model for $y_i$. For example, in image applications, after training a normalizing flow network, one could generate a new Gaussian vector $z_{i'}$ and push it forward through the trained network, with the transformed $y_{i'}$ often looking as if it were a real photo. Since there is only one neural network involved (despite a large number of parameters within it), the computation enjoys high efficiency through stochastic gradient descent. Its expressiveness as a generative model, tractability of the target density $f(y_i)$ and good computing performance make the normalizing flow a compelling alternative to mixture models for high-dimensional density estimation provided the training data set is large enough.

On the other hand, since one can inversely obtain $z_i$ as a deterministic transform of $y_i$, a number of interesting directions of exploration for $z_i$ arise. One of them is whether a more interpretable modeling structure for $z_i$ can be used, other than just being drawn from an independent Gaussian distribution. Early examples includes using a mixture of Gaussian distribution  \citep{izmailov2020semi}, or a mixture of subspace structure \citep{peng2020deep} to name a select few. Although some interpretable results, including improved clustering accuracy were reported, it was later discovered that most of the improved results were largely due to the specific pre-processing of the reported data sets \citep{haeffele2020critique}, instead of the selected distribution for $z_i$. This cautionary tale serves as a good warning, that it is quite difficult---if not impossible---to rely on unsupervised normalizing flow (that is, using $y_i$ alone) to find structure in the latent $z_i$. Naturally, this motivates us to consider external information from $x_i$, and create a normalizing flow-based conditional density estimator.

The focus is on the CDE problem involving a high-dimensional $y_i$ and low-dimensional $x_i$; for example, $x_i$ could correspond to labels, or a continuous vector providing context information for the observed $y_i$.
Specifically, we use the normalizing flow to form an invertible transform of data $y_i$ and an ``augmented posterior (AP)'' based $z_i$ that comprises of two components $[z_{P,i}, z_{N,i}]$; $z_{P,i}$ is a low-dimensional subvector that forms a joint distribution with $x_i$, such as via simple logistic/linear regression likelihood $f(x_i\mid z_{P,i})$.
Effectively, the distribution of $[z_{P,i}, z_{N,i}]$ is the posterior distribution of $(z_{P,i}\mid x_i)$ augmented by independent Gaussian $z_{N,i}$. Note that there has been some recent work on normalizing flow based CDE, e.g., \cite{papamakarios2017masked}. However, the proposed approach enjoys several unique advantages: first, it produces a supervised dimension reduction in those $z_{P,i}$ within the CDE framework; second, it requires only
simple modification of the common normalizing flow networks, so it is very easy to implement; last, it produces a single latent variable $z_i$ for each data point $y_i$, and hence the unconditional density calculation does not involve summation or integration over the space of the predictor variable. We will illustrate these advantages as well as the outperformed density estimation results in this article.

\section{Preliminary: Generative Models based on Normalizing Flows}

This section provides some preliminary on the normalizing flow neural networks. Suppose there is a one-to-one and differentiable almost everywhere mapping $T_\theta: \mathbb{R}^p \to \mathbb{R}^p$ (with $\theta$ the parameters within it), that can transform random variable $y$ into another latent continuous $z\in\mathbb{R}^p$ following a simple distribution $f_z$.
With a slight abuse of notation, $f_.$ is used to represent both a  density and a probability function. After a change of variable, for $i=1,\ldots,n$:
\[
f_y(y_i) = f_z [ T_{\theta}(y_i) ] |\nabla_y T_{\theta}( y_i)|,
\]
with $|\nabla_y T_{\theta}(\cdot)|$ being the determinant of the Jacobian matrix with the gradient taken with respect to $y_i$. To estimate  $T_{\theta}$, one typically minimizes the Kullback--Leibler (KL) divergence between the target distribution $f_y^*(y)$ and the normalizing flow-based $f_y(y)$, which is
\be
\text{KL}[ f_y^*(y) , f_z [ T_{\theta}(y) ] |\nabla_y T_{\theta}( y)|] 
\approx    \frac{1}{n} \sum_{i=1}^n
\bigl\{
-\log f_z [ T_{\theta}(y_i) ] |\nabla_y T_{\theta}( y_i)|
\bigr\}+  \text{constant},
\ee
where the right hand side is the empirical KL, equaling to the loss function to be minimized over $\theta$.

To flexibly parameterize the transform while maintaining invertibility, one uses a special multilayer neural network (``invertible neural network'') with $T=T_1\circ T_2 \circ \cdots \circ T_m$, and each $T_k:\mathbb{R}^p\to \mathbb{R}^p$ is a layer of an invertible transform of relatively simple operations; for example, in RealNVP \citep{dinh2016density}, the input of $T_k$ is equally partitioned to $r=[r_A,r_B]$, and $T_k([r_A,r_B])= [r_A, r_B \odot s(r_A) + l(r_A) ]$, with $s$ and $l$ functions that produce the location-scale change to $r_B$. Then in the next layer, one alternates by using $T_{k+1}([r_A,r_B])= [r_A \odot s(r_B) + l(r_B), r_B ]$. Other types of neural networks include autoregressive flows and residual flows. We refer readers to \cite{papamakarios2021normalizing} as a review for all types of flows employed. Besides invertibility, these neural networks are also carefully designed so that the term of the determinant of the Jacobian and the inverse mapping of the neural networks can be computed at low cost.

Obtaining an approximate solution of $T_\theta$ via an invertible neural network leads to a transport map $T_{\hat\theta}$, that gives a density estimator for the data  $\hat f_y(y_i) = f_z [ T_{\hat\theta}(y_i) ] |\nabla_y T_{\hat\theta}( y_i)|$. This method is commonly referred to as ``normalizing flow'', since one often assigns a standard independent Gaussian distribution to $z\sim \text{N}(0,I_p)$ for simplicity, which is often called the base distribution. As a generative model, one can sample $z_{i'}\sim \text{N}(0,I_p)$, and then $y_{i'} = T^{-1}_{\hat\theta}(z_{i'})$ produces new generated data from the estimated distribution $f_y$. 

\section{Method}

The section begins by introducing notation used in the sequel. Let $y_i\in \mathbb{R}^p$ be a continuous response with distribution $f_y$, and $x_i\in \mathcal X \subseteq \mathbb{R}^m$ be the corresponding predictor variable (could be discrete, continuous, or a mix of both) drawn from another distribution $f_x$. We focuses on the case of large $p$ and small $m$.

\subsection{Augmented Posterior for CDE}\label{subsec:augpos}

To enable CDE for $f_{y\mid x}(y_i\mid x_i)$, as well as making the latent variable $z_i$ more interpretable, we consider a joint distribution between $z$ and $x$,
\begin{equation}\label{eq:joint_z_x}
	f_{z,x}(z_i,x_i) =    f_{z_N}(z_{N,i})   f_{z_P}(z_{P,i}) f_{x\mid z}(x_i\mid z_{P,i};\beta),
\end{equation}
independently for $i=1,\ldots,n$,
where $z_i=[z_{P,i}, z_{N,i}]$. The first component $z_{P,i}\in \mathbb{R}^d$ is a low-dimensional subvector and is used in a predictive model for $x_i$, whereas the second component
$z_{N,i}\in \mathbb{R}^{p-d}$ is a high-dimensional subvector unrelated to $x_i$, and $\beta$ is viewed as a non-random parameter that will be estimated. If one considers $f_{x\mid z}$ to be the likelihood for $x$, and $f_z$ to be the prior distribution for $z$, then it is not hard to see that,
$$
f_{z\mid x}(z_i\mid x_i)= f_{z_N}(z_{N,i})   \frac{f_{z_P}(z_{P,i}) f_{x\mid z}(x_i\mid z_{P,i};\beta)}{
	\int_{\mathbb{R}^d} f_{z_P}(t) f_{x\mid z}(x_i\mid t;\beta) \textup{d}t
},
$$
where the second part is the posterior of $z_P$, $f_{z_P\mid x}(z_{P,i}\mid x_i;\beta)$,  and $f_{z_N}(z_{N,i})$ is an independent random variable that augments this posterior, to make $z_i$ match the dimension of $y_i$. Therefore, each $z_i=(z_{P,i},z_{N,i})$ is referred to as a sample point from an \textit{augmented posterior}.

\begin{figure}[htp]
	\centering
	\includegraphics[width=0.5\linewidth]{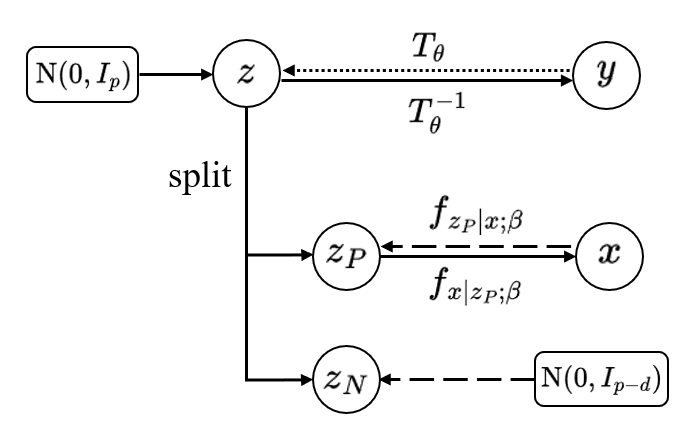}
	\caption[The diagram of the architecture of AP-CDE.]{The diagram of the architecture of AP-CDE. The solid lines show the generative process for the data $(y,x)$. The dashed lines show how to generate a new latent variable $z$ from the augmented posterior. \label{fig:diag}}
\end{figure}

\cref{fig:diag} shows the architecture of the proposed model. Note that for simplicity, we retain the standard independent Gaussian density for $f_{z_N}$ and $f_{z_P}$ as in a typical normalizing flow, while employing a generalized linear model for $f_{x\mid z}$. For example, if $x_i\in \mathbb{R}^m$ is continuous, then one can use $x_i = \beta_0 + \beta_1 z_i+ \epsilon_i$, with $\beta_0\in \mathbb{R}^m$, $\beta_1$ an $m\times d$ matrix, and $\epsilon_i \stackrel{iid} \sim \text{N}[ \vec 0, \text{diag} (\beta_2)]$ with $\beta_2$ positive vector. For each univariate discrete $x_{i,j}\in \{1,\ldots, K\}$, one can use a multinomial logistic regression in $f_{x\mid z}$, $p(x_{i,j}=k\mid z_{P,i}) \propto \exp(\beta_{0,k }+\beta_k' z_{P,i})$, with $\beta_{0,k}\in\mathbb{R}$, $\beta_{k}\in \mathbb{R}^d$ for $k=1,\ldots, K-1$, and  $\beta_K$ and $\beta_{0,K}$ fixed to 0-value as common in logistic regression.
In general, the conditional probability function \(f_{x\mid z}\) could be proportional to the original function to the power \(\lambda\) for the purpose of regularization. This is commonly used in robust Bayesian models \citep{grunwald2017inconsistency} and hybrid deep generative models \citep{nalisnick2019hybrid}. In the proposed model, the normalizing constant is merged into the denominator of the posterior $f_{z_P\mid x}$ naturally, and the model is still a generative one.

In the case of $x_i$ having both continuous and (potentially more than one) discrete elements, $x_i=[x_{A,i}, x_{B,i}]$, one can partition $z_{P,i}= [z_{P_A,i}, z_{P_B,i}]$, and use a separable likelihood function $f_{x\mid z_P}(x_i\mid z_{P,i}) = f_{x_A\mid z_{P_A}} (x_{A,i}\mid z_{P_A,i}) f_{x_B\mid z_{P_B}}(x_{B,i}\mid z_{P_B,i})$---where the first term on the right models the continuous part, and the second term does the discrete part. This separation in $z_P$ is motivated by applications under consideration, in which commonly the discrete part corresponds to the class label of an image, whereas the continuous part to other conditions (such as lighting) that are unrelated to the labeling information.

Next, if there is an invertible mapping $T_\theta$ that connects $z_i$ and $y_i$, by applying change-of-variable, one has
$$
f_{y\mid x}(y_i\mid x_i) = f_{z\mid x}[T_\theta(y_i) \mid x_i;\beta]   |\nabla_y T_{\theta}( y_i)|.
$$

To estimate this mapping $T_\theta$ as well as the parameter $\beta$ in the predictive model of $x_i$, one minimizes the empirical KL divergence between the target distribution $f_{y\mid x}^*(y\mid x)$ and the $f_{y\mid x}(y\mid x)$ in the above, leading to:
\begin{equation}\label{eq:KL_AP}
	\begin{aligned}
		\min_{\theta,\beta} \frac{1}{n}\sum_{i=1}^n \bigg\{ & -\log
		f_{z}[ T_\theta(y_i) ]     |\nabla_y T_{\theta}( y_i)| \\
		&  -\log f_{x\mid z}[x_i\mid T^P_\theta(y_i);\beta] 
		+ \log {
			\int_{\mathbb{R}^d} f_{z_P}(t) f_{x\mid z}(x_i\mid t;\beta) \textup{d}t
		}
		\bigg\},
	\end{aligned}
\end{equation}
where $T^P_\theta(z_i)$ [or, $T^N_\theta(z_i)$] means taking the subvector (corresponding to $z_P$, or $z_N$) from the output of $T_\theta(z_i)$. Although the last integral above is often intractable, as the stochastic gradient descent technique (the commonly used optimization algorithm in normalizing flow) only requires an approximate gradient via taking a random  subset of size $n_b$ instead of $n$, one can replace the last term via a Monte Carlo estimate, $\log \bigl[ \sum_{l=1}^M f_{x\mid z}(x_i\mid t_l;\beta)/M\bigr]$, with each $t_l\stackrel{iid}\sim f_{z_P}$ [in this article, $\text{N}(\vec 0, I_d)$]. The $n_b$ integrals to be estimated can share those $M$ random samples $t_l$'s across $i$ in the subset of size $n_b$, and hence the estimation is inexpensive.

After the KL divergence is minimized, a conditional density estimator is obtained by
$$\hat f_{y\mid x}(y_i\mid x_i)=f_{z\mid x}\bigl[T_{\hat\theta}(y_i) \mid x_i;\hat\beta\bigr]   |\nabla_y T_{\hat\theta}( y_i)|.$$
Further, to calculate the marginal density of $y$ for a new data point for which only the response $y_{i}$ is available, one can simply marginalize over $x_i$ in \cref{eq:joint_z_x}, and obtain
\[
f_y(y_i)=f_{z_N}\bigl[T^N_{\hat\theta}(y_{i})\bigr] f_{z_P}\bigl[T^P_{\hat\theta}(y_{i})\bigr]\bigl|\nabla_y T_{\hat\theta}( y_i)\bigr|
.\]
Therefore, after the optimization, both conditional and marginal density estimation can be accomplished in a computational efficient manner.

\subsection{Supervised Dimension Reduction and Validation}\label{subsec:dimreduction}
Besides conditional density estimates, another advantage of using the augmented posterior is that it produces a low-dimensional representation $z_{P,i}$ that is related to the variations in $x_i$. Indeed, after the optimization step concludes, one obtains a computational form that produces low-dimensional $z_{P,i}=T^P_{\hat\theta} (y_i)$, and $T^P_{\hat\theta}$ is estimated under supervising information from $x$.

On the other hand, one may further wonder if $z_{P,i}$ has captured all the useful information that connects $y_i$ and $x_i$. To formalize, if the true data generating mechanism for $(y_i,x_i)$ is indeed based on $z_i\sim f_z$, $y_i= T^{-1}_{\theta}(z_i)$, $x_i \sim f_{x\mid z_P}$, then one would have the following sufficient dimension reduction outcome:
$$x \indep y \mid T^P_\theta(y),$$ as the ideal result.

There are ways to test conditional independence in classical linear models \citep{cook2005sufficient} and non-linear low dimensional models \citep{su2007consistent,su2008nonparametric}. However, the combination of nonlinearity and high dimensionality poses significant challenges. Fortunately, for high-dimensional data with $y_i$ and $x_i$, this can be validated via synthesizing new data and predicting $x_i$ via another neural network $G$, independently trained with $(y_i,x_i)$'s.

Specifically, for each $z_i=T_{\hat\theta}(y_i)$ produced, one fixes $z_{P,i}$ while replacing $z_{N,i}$ with an independently sampled $\tilde z_{N,i,j}\sim \text{N}(0, I_{p-d})$ for $j=1,\ldots,J$. Then synthesized $\tilde{y}_{i,j}= T_{\hat\theta}^{-1} [z_{P,i}, \tilde z_{N,i,j}]$ is obtained. One predicts the $x_{i}$ using $\tilde{y}_{i,j}$ via the separately trained network $G$, and observes if each predicted $\hat x_{i,j}$ differ from the observed $x_i$---in the ideal case, $\hat x_{i,j}$ should not differ much from $x_i$ (since $z_{P,i}$ should contain most of the predictive information about $x_i$), and the conditional independence can be quantified by the error rate.

\subsection{Parameterization Details}
In this article, $T_\theta$ is parameterized using the state-of-art normalizing flow Glow \citep{kingma2018glow}. We choose Glow in the experiments for its high accuracy on the density estimation and the ease of implementation. It employs a multi-scale architecture \citep{dinh2016density}, which contains $L$ levels. After each level, half of the dimensions of latent $z_i$ are immediately modeled as Gaussians, while the remaining half are further transformed by the flows. This significantly improves the computation efficiency. Each level consists of $K$ (depth) steps of flow that share an identical structure which contains an activation normalization, an invertible $1\times 1$ convolution and an affine coupling layer. In this article, we use the additive coupling layer as a special case of the affine coupling layer, in which the number of channels of the hidden layers (convolutional neural networks) is set to be $512$.

Under this kind of multi-scale architecture, the following important problem is how to choose the way of splitting $z_i$ into $[z_{P,i},z_{N,i}]$. First, to make the Monte Carlo estimation of the integral in \cref{eq:KL_AP} accurate, efficient and stable, one does not expect $d$, the dimension of $z_{P,i}$, too large. Second, since the outputs from the later layers experience more transforms compared to the ones from the earlier layers, choosing dimensions of $z_{P,i}$ from the relatively later layers will improve the model performance. This is illustrated in the numerical experiments.

\section{Numerical Experiments}\label{section:expr}

The following set of numerical experiments demonstrates the AP-CDE's advantages for density estimation and dimensional reduction. The following competing models are used to compare its performance: (i) the Glow normalizing flow based on independent Gaussian $z\sim \t{N}(\vec 0,I_p)$, without using any information from $x$; (ii) a modified Glow based on a Gaussian mixture (named Glow-Mix), $z_i\sim \sum_{k=1}^K w_k \t{N}  (\mu_k,I_p)$, with $K$ set to the ground-truth number of classes in the data; and (iii) a CDE extended from Glow (named Glow-CDE) by adding $x_i$ as an input in each additive coupling layer \cite[Section 3.4]{papamakarios2017masked}. When the predictor variable $x_i$ is discrete, e.g., class label, the comparison is also made with (iv) a naive CDE model based on Glow where for each value $x'$ of $x_i$ a Glow is trained on those data $y_i$ with $x_i=x'$ (named Glow-NCDE). For a fair comparison, we set the four competitors to have the same numbers of levels $L$ and the same depth $K$, as in the AP-CDE model. The first two models perform unconditional density estimation with producing latent variables, whereas the last two models are conditional. Unlike the proposed model, the two CDE models compared do not have a unique corresponding latent variable for a new data. Note that the fourth model is inefficient because one has to train models of number of classes of $x$.

We use the Adam optimizer \citep{kingma2014adam} provided in the PyTorch framework to train all models, with a mini batch size of $n_b=64$ and learning rate at $0.0005$ for all models. All models are trained for $200$ epochs, where the optimizer goes through the whole training dataset exactly once in each epoch. For the first $10$ epochs as warm-up, the learning rate linearly increases to $0.0005$ after each batch training. Then the learning rate goes down to $10^{-4}$ using cosine annealing schedule \citep{loshchilov2016sgdr}. We monitor the loss function of all models and ensure that convergence is achieved by all models. Finally, $M=1000$ is set in the Monte Carlo estimator for the integral in \cref{eq:KL_AP}.

The experiments are based on the following two datasets: the FashionMNIST one of the fashion products \citep{xiao2017fashion} and the Extended Yale Face B one of face images \citep{lee2005acquiring}. These images have a single color channel, containing pixel values $\{0,\dots,255\}$. We follow \citet{papamakarios2017masked, dinh2016density} to dequantize the pixel values by adding standard uniform noise onto every pixel and scaling the values to \((0,1)\) by dividing $256$. Additional numerical experiments are shown in the \cref{sec:mnist}.

\subsection{FashionMNIST Images of Fashion Products}

The FashionMNIST data is used to illustrate the CDE when $x_i$ is discrete. This dataset contains $70,000$ processed images of fashion products, each having $28\times 28$ pixels. Among them, $60,000$ is used for training purposes and the remaining is used for testing. Each image $y_i$ is associated with a discrete label $x_i$ with values from $0$ to $9$ recording the ground-truth fashion products, including T-shirt, sandal, bag, etc. Each image is padded to dimension $32\times 32$ by adding 2 more dimensions of $0$ in each of four direction, and then extend each image to $3$ channels by repeating the image in each of channel.

For the Glow model, levels $L=3$ and depth $K=32$ are set.
For the sub-model \(f_{x\mid z}\) in AP-CDE, we use the likelihood function of the multinomial logistic regression, as stated in \cref{subsec:augpos}. Since the labels are relatively balanced across classes, to improve interpretation on the latent $z_{P,i}$'s, all intercept terms $\beta_{0,k}$'s are set to be zeros. We compare several choices of the dimensions of $z_{P,i}$. To be clear, in the multi-scale architecture, when the data $y_i$ has dimension $3\times 32 \times 32$, the output $z_i^{(1)}$ from the first level has dimension $6\times 16\times 16$, the output $z_i^{(2)}$ from the second level has dimension $12\times 8\times 8$, while the final output $z_i^{(3)}$ has dimension $48\times 4\times 4$. The following choices of $z_{P,i}$ are compared: (1) $z^{(3)}_{1:2,1,1}$; (2) $z^{(2)}_{1:2,1,1}$; (3) $z^{(1)}_{1:2,1,1}$; (4) $z^{(3)}_{1:16,1,1}$; (5) $z^{(2)}_{1:4,1:2,1:2}$; (6) $z^{(1)}_{1:4,1:2,1:2}$; (7) $z^{(3)}_{1:48,1:2,1:2}$; (8) $z^{(2)}_{1:12,1:4,1:4}$; and (9) $z^{(1)}_{1:3,1:8,1:8}$.

\begin{figure}[ht]
	\centering
	\captionsetup[subfigure]{justification=centering}
	\begin{subfigure}[b]{0.32\linewidth}
		\centering
		\includegraphics[width=1\linewidth]{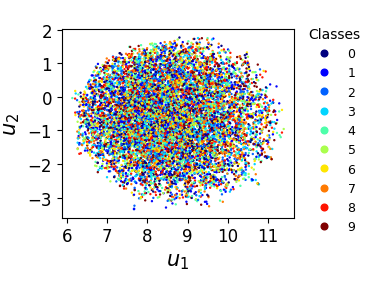}
		\caption{Glow}
	\end{subfigure}
	\begin{subfigure}[b]{0.32\linewidth}
		\centering
		\includegraphics[width=1\linewidth]{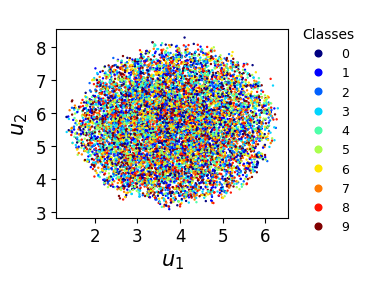}
		\caption{Glow-Mix}
	\end{subfigure}
	\begin{subfigure}[b]{0.32\linewidth}
		\centering
		\includegraphics[width=1\linewidth]{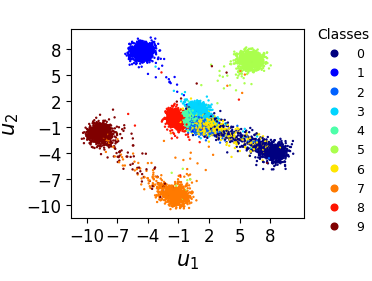}
		\caption{AP-CDE}
	\end{subfigure}
	\caption[Latent representations estimated by the three models applied on the FashionMNIST training set.]{Latent representations estimated by the three models applied on the FashionMNIST training set. For the Glow and the Glow-Mix models, UMAP is used to reduce the dimensions to $2$. \label{fig:fmnist}}
\end{figure}

\cref{fig:fmnist} plots the latent representations produced by Glow, Glow-Mix and AP-CDE with the choice (1) for the $z_{P,i}$. Recall that for Glow and Glow-Mix, the latent $z_i$ has the same dimensionality as the images, UMAP \citep{mcinnes2018umap} is used to reduce the dimension and plot its output in 2D. For AP-CDE, the plot of the latent $z_{P,i}$ is provided. As expected, the latent variable produced by Glow follows a simple spherical Gaussian, and thus is not interpretable. Somewhat surprising, Glow-Mix does not produce a meaningful result either, despite using a mixture of $10$ Gaussians (that corresponds to the true number of classes)---instead, the Glow-Mix model converges to only one component containing a mixture of $z_i$'s from all classes. This negative finding is in accordance to an early critique on clustering with deep neural networks \citep{haeffele2020critique}, where it was reported that in an unsupervised setting, imposing a modeling structure on the latent variable (such as a mixture of Gaussians) does not lead to a clear separation of data from different classes.  Using AP-CDE and the supervising label information form $x_i$, we obtain a good separation of the ten products based on the low-dimensional representation $z_{P,i}\in\mathbb{R}^2$ for the model $z^{(3)}_{1:2,1,1}$ (shown in \cref{fig:fmnist}(c)).

\begin{table}[ht!]
	\centering
	\caption[The averages of bits per dimension on the training and the testing sets of FashionMNIST for all models.]{The averages of bits per dimension on the training and the testing sets of FashionMNIST for all models. Lower BPM means higher density.}
	\label{tab:den}
	\begin{tabular}{lrrrrr} \toprule
		{Models}    &{Glow} & {Glow-Mix} & {Glow-CDE} & {Glow-NCDE} & {AP-CDE}\\
		\midrule
		Training set    & 1.02  & 1.04 & 1.03  &  1.31  & 1.02 \\
		Testing set     & 1.03  & 1.05 & 1.04  &  1.32  & 1.03 \\
		\bottomrule
	\end{tabular}
\end{table}

\cref{tab:den} depicts the average bits per dimension (BPM) on the training and the testing sets for all models. Here the BPM is the negative log-densities divided by the number of dimensions, which is broadly used in the literature because it has similar scale for different resolution of images.
For AP-CDE model, we choose the best model considering the error rate and the density estimation performance among the all choices of $z_{P,i}$. The model of using $z^{(3)}_{1:48,1:2,1:2}$ is chosen.  Clearly, for this dataset, AP-CDE, likely due to a better group-wise concentration, produces overall higher (or equal) marginal densities compared to its competitors. 

\begin{table}[ht!]
	\centering
	\caption{The averages of bits per dimension on the training and the testing sets and the classification error rates on the testing sets for all AP-CDE models on the FashionMNIST data.}
	\label{tab:error}
	\begin{tabular}{lccccc}
		\toprule
		& {$z^{(3)}_{1:2,1,1}$} & {$z^{(2)}_{1:2,1,1}$} & {$z^{(1)}_{1:2,1,1}$} & {$z^{(3)}_{1:16,1,1}$} & {$z^{(2)}_{1:4,1:2,1:2}$} \\
		\midrule
		BPM (Training set) & 1.02 & 1.03 & 1.01 & 1.05 & 1.06 \\
		BPM (Testing set)  & 1.03 & 1.03 & 1.02 & 1.06 & 1.06 \\
		Error rate (\%)    & 50.96 & 60.53 & 89.69 & 6.89 & 7.48 \\
		\bottomrule
	\end{tabular}
	
	\vspace{1em}

	\begin{tabular}{lcccc}
		\toprule
		& {$z^{(1)}_{1:4,1:2,1:2}$} & {$z^{(3)}_{1:48,1:2,1:2}$} & {$z^{(2)}_{1:12,1:4,1:4}$} & {$z^{(1)}_{1:3,1:8,1:8}$} \\
		\midrule
		BPM (Training set) & 1.02 & 1.02 & 1.04 & 1.01 \\
		BPM (Testing set)  & 1.02 & 1.03 & 1.05 & 1.02 \\
		Error rate (\%)    & 80.33 & 6.46 & 7.26 & 89.9 \\
		\bottomrule
	\end{tabular}
\end{table}

\cref{tab:error} depicts the average bits per dimension on the training and the testing sets for all AP-CDE models as well as the classification error rates on the testing set, where one obtains $z_{P,i}$ and predict the label via the trained logistic regression model using \(\arg\max_{k\in\{0,\ldots,9\}} f_{x\mid z_P}(k\mid T_{\hat\theta}^P(y_i))\). It shows that when one chooses more dimensions of the same level for the $z_{P,i}$, the classification error rate will be lower, while the density estimation performances do not vary significantly. Moreover, even when the $z_{P,i}$ has the same number of dimensions, the higher level can provide more accurate classification. This bunch of experiments is an important guidance on how to choose the dimensions for $z_{P,i}$---the whole output of the last level is always not a bad choice. 

Empirically it shows that the low-dimensional representation $z_{P,i}$ contains almost all the information to separate the different classes when the dimensions of $z_{P,i}$ are well chosen. For the model $z^{(3)}_{1:48,1:2,1:2}$, as described in \cref{subsec:dimreduction}, if one fixes $z_{P,i}$, but replaces $z_{N,i}$ with independently sampled realizations from a Gaussian distribution, and then through $T^{-1}_{\hat \theta}$ one can obtain $10$ new images for each original observation $i$. We then employ the ResNet101 \citep{he2016deep} (a convolutional neural network separately trained on the training set) to classify these artificially generated images, and find that $95.26\%$ of them are still classified to the same class label as the $y_i$'s. Hence, it is concluded that the $z_{N,i}$ largely corresponds to within-class variation, whereas $z_{P,i}$ captures between-class variation.

\begin{figure}[ht!]
	\centering
	\captionsetup[subfigure]{justification=centering}
	\begin{subfigure}[t]{0.35\linewidth}
		\centering
		\includegraphics[width=1\linewidth]{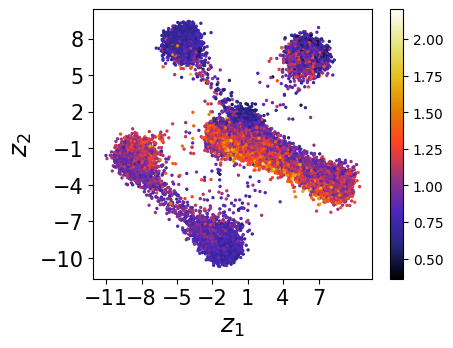}
		\caption{Training set.}
	\end{subfigure}
	\begin{subfigure}[t]{0.35\linewidth}
		\centering
		\includegraphics[width=1\linewidth]{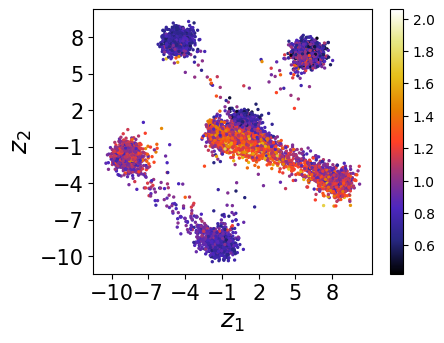}
		\caption{Testing set.}
	\end{subfigure}
	\caption[The first two dimensions of the latent variables from AP-CDE model.]{The first two dimensions of the latent variables from AP-CDE model $z^{(3)}_{1:2,1,1}$ on FashionMNIST, colored by the estimated densities in the scale of BPM. \label{fig:fmnist_den}}
\end{figure}

\begin{figure}[ht!]
	\centering
	\includegraphics[width=0.75\linewidth]{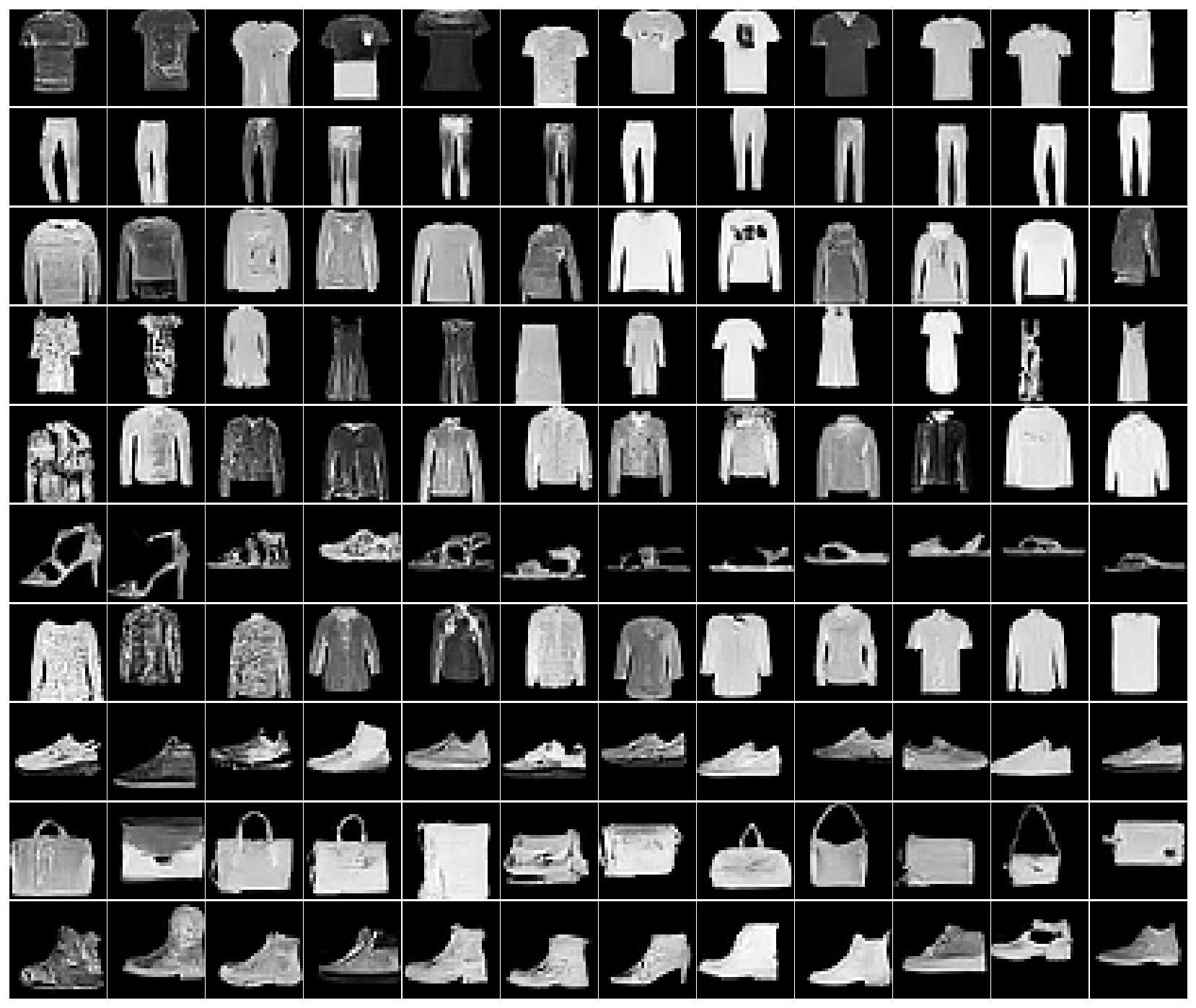}
	\caption[Sample images from the AP-CDE model trained by FashionMNIST data.]{Sample images from the AP-CDE model $z^{(3)}_{1:48,1:2,1:2}$ trained by FashionMNIST data, with each row sorted in the increasing order of estimated densities. \label{fig:gene_fmnist}}
\end{figure}

Further, we color the latent $z_{P,i}$ using the magnitude of the density (\cref{fig:fmnist_den}). In the result, those points with relatively low density values tend to correspond to images of low quality or higher ambiguity regarding the product class. To show this, we plot a few sampled fashion products in \cref{fig:gene_fmnist} and sort each row by the density value in increasing order. It can be seen that the images on the left tend to be harder to assign to a class, compared to the ones on the right.

\subsection{Human Face Photos}

To illustrate the AP-CDE with continuous and mixed-type $x_i$, experiments are also run on the Yale face dataset. There are $2,414$ face photos, each containing a single color channel with a $168\times 192$ pixel resolution.  The images are resized to $28\times 32$ to reduce computation cost, while maintaining clarity of the photos. The images come from $38$ people, and this information is used as a discrete class variable
$x_{A,i}$. Further, the photos were taken under different light conditions, recorded as azimuth angle and elevation. This information is used as two continuous variables $x_{B,i}$ and $x_{C,i}$. 

For the Glow model, the parameters are set to $L=3$ and $K=32$. 
A logistic regression likelihood $f_{x_A\mid z_{P_A}}$  is used for the discrete $x_{A}$, that depends on $3$ dimensions of $z^{(3)}_{1:3,1,1}$. Here, $z^{(3)}$ is the output from the third level. Linear regression $x_{B,i} = \beta^B_0 + \beta^B_1 z_{P_B,i}+ \epsilon^B_i$, $x_{C,i} = \beta^C_0 + \beta^C_1 z_{P_C,i}+ \epsilon^C_i$ are employed for the other two covariates, where both $z_{P_B,i}=z^{(3)}_{4,1,1}$ and $z_{P_C,i}=z^{(3)}_{5,1,1}$ are one dimensional. Assume $\epsilon^B_i \stackrel{iid} \sim \text{N}(0, 0.01)$ and $\epsilon^C_i \stackrel{iid} \sim \text{N}(0, 0.01)$; note the low value of the variance selected, which forces higher correlation between $(x_{A},z_{P_A})$ and $(x_{B},z_{P_B})$. In this case, the last integral in \cref{eq:KL_AP} has closed form because the integrand is the product of two Gaussian densities. The integral is $1/\sqrt{2\pi(\beta_2^2+ \beta_1^2 )} \exp\{-(x_i-\beta_0)^2/[2(\beta_2^2+\beta_1^2)]\}$.

\begin{figure}[htp]
	\centering
	\captionsetup[subfigure]{justification=centering}
	\begin{subfigure}[c]{0.35\linewidth}
		\centering
		\includegraphics[width=1\linewidth]{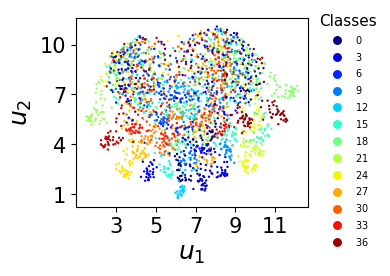}
		\caption{Glow}
	\end{subfigure}
	\begin{subfigure}[c]{0.35\linewidth}
		\centering
		\includegraphics[width=1\linewidth]{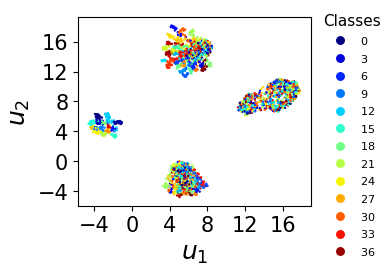}
		\caption{Glow-Mix}
	\end{subfigure}
	\begin{subfigure}[c]{0.25\linewidth}
		\centering
		\includegraphics[width=1\linewidth]{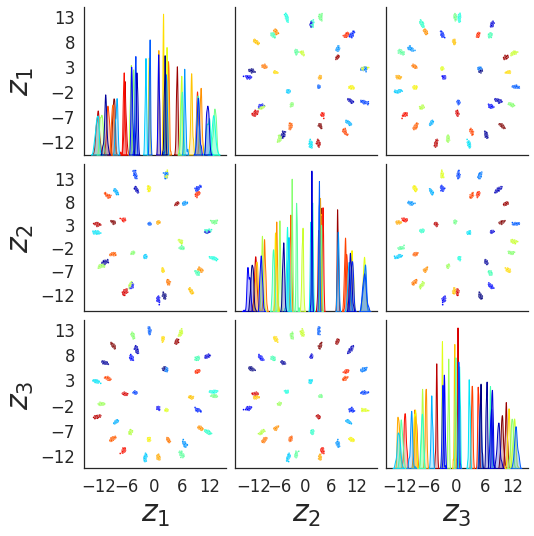}
		\caption{AP-CDE}
	\end{subfigure}
	\caption[The latent representations estimated from the Yale face data.]{The latent representations estimated from the Yale face data. The UMAP is used to reduce the dimensions to 2 for Glow and Glow-Mix. For the AP-CDE model, all the dimensions of $z_P$ are shown in pairs plot. \label{fig:yaleb}}
\end{figure}

Competing methods include Glow and Glow-Mix.  As can be seen in \cref{fig:yaleb}, AP-CDE leads to a clear separation of latent representations due to the use of label information, whereas unsupervised Glow and Glow-Mix fail to do so. Further, the Glow model does not produce a group-mixed sphere as did in the FashionMNIST experiment. The Glow-Mix model produces four clusters in the latent space but none of them represents some class of people.

\begin{figure}[ht!]
	\centering
	\includegraphics[width=.7\linewidth]{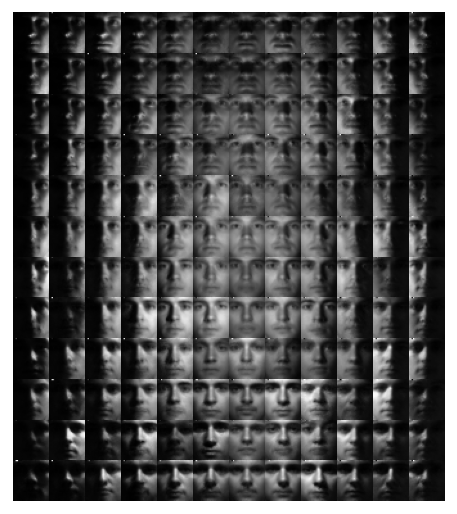}
	\caption[Synthesized face photos with gradually changing light azimuth angles and elevations.]{Synthesized face photos with gradually changing light azimuth angles (left to right) and elevations (top to bottom).  \label{fig:gene_yaleb}}
\end{figure}

To show the expressiveness of AP-CDE as a generative model, the following procedure is used to synthesize new artificial images. One selects a range for Azimuth angles ($-100$ to $100$) and a range for elevations ($-60$ to $60$) (with interpolation), and form a $12\times 12$ grid. For each grid cell, one draws a $Z_{P_A,i}$ that corresponds to a person's identity, from the empirical posterior distribution from the AP-CDE estimates. Further, one randomly draws the $Z_{N,i}$ component from a $\text{N}(\vec 0,I)$ distribution.
With this choice of $\tilde z_i$, we synthesize new images for the $i$-th subject $\tilde y_i = T_{\hat \theta}^{-1}(\tilde z_i )$. As shown in \cref{fig:gene_yaleb}, there is a clear trend of change in the lighting conditions, caused by the changing values of $(x_{B,i},x_{C,i})$.

\section{Data Application}

In the application study, we use 18001 images of leaves from strawberry plants, which either are healthy or have one of the three types of diseases: powdery mildew, anthracnose and fusarium wilt. This forms the labels of the four classes, which can be treated as the predictor variable $x$. The AP-CDE model is expected to conditionally estimate the densities of the images as well as do a supervised dimensional reduction.

Considering that the backgrounds of the images cause overfitting problem because the similarities between backgrounds rather than between the characteristics of the diseases take account for the major contribution on the distinction of the several groups \citep{saikawa2019aop}, we use the method mentioned in \cite{saikawa2019aop} to remove the backgrounds of all images. This segmentation also enhanced the quality of density estimation by reducing the redundant information. The images are resized to $128 \times 128$ resolution. To improve visual quality in generating samples, we follow \cite{kingma2018glow} to use 5-bit images. Then they dequantize the pixel values as stated in the \cref{section:expr}. 
The data is split into training set which includes 16000 images and the testing set which includes 2001 images. 

For the parameters of Glow, the parameters are set to $L=6$ and $K=32$.  
For the conditional likelihood of $x\mid z_P$, again, we use $f_{x\mid z}\propto g^{\lambda}_{x\mid z}$, where $g$ is the likelihood function of the multinomial logistic regression through the origin and $\lambda=1000$. The subvector $z_{P,i}$ is chosen to be the output of the last level, which has dimension $384 \times 2 \times 2$. 

\begin{table}[ht!]
	\centering
	\caption{The averages bits per dimension on the training and the testing sets of the leaves of strawberry plants data for all models.}
	\label{tab:straw_den}
	\begin{tabular}{lrrrrr} \toprule
		{Models}    &{Glow} & {Glow-Mix} & {Glow-CDE} & {Glow-NCDE} & {AP-CDE}\\
		\midrule	
		Training set  & 4.16 & 4.25 & 4.19 & 4.18 & 4.16\\
		Testing set & 4.17 & 4.26 & 4.22 & 4.18 & 4.17\\
		\bottomrule
	\end{tabular}
\end{table}

\begin{figure}[H]
	\centering
	\captionsetup[subfigure]{justification=centering}
	\begin{subfigure}[t]{0.4\linewidth}
		\centering
		\includegraphics[width=1\linewidth]{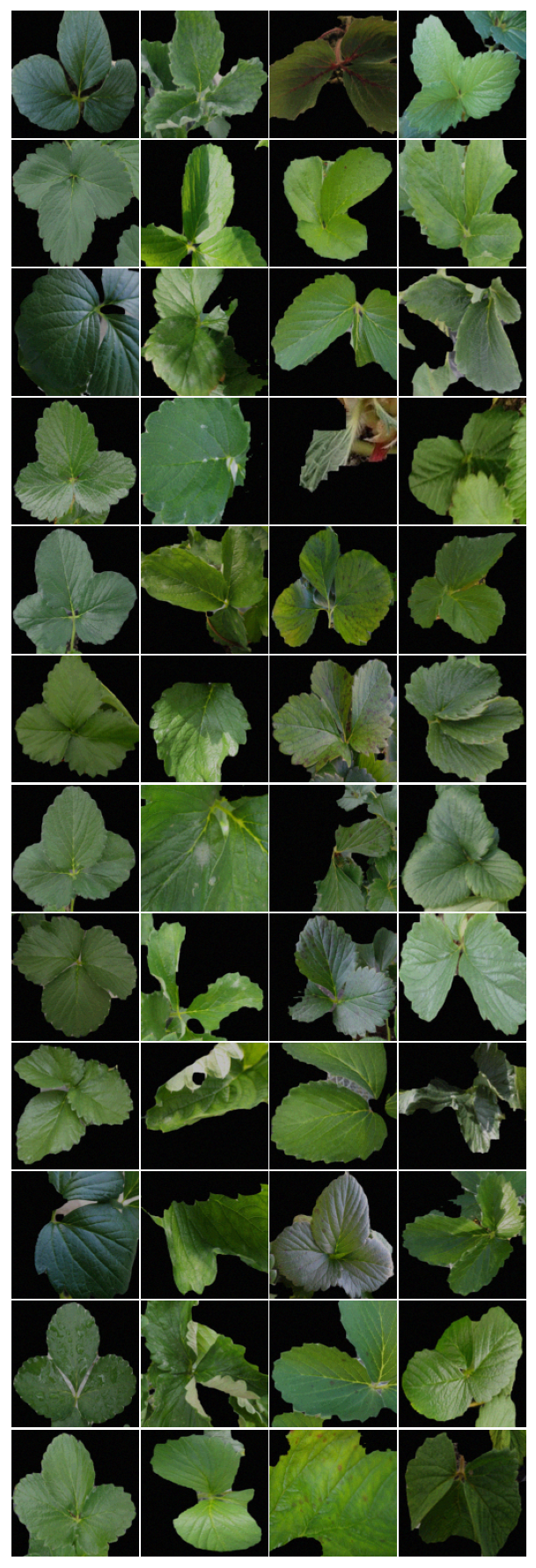}
		\caption{Real images}
	\end{subfigure}
	\begin{subfigure}[t]{0.4\linewidth}
		\centering
		\includegraphics[width=1\linewidth]{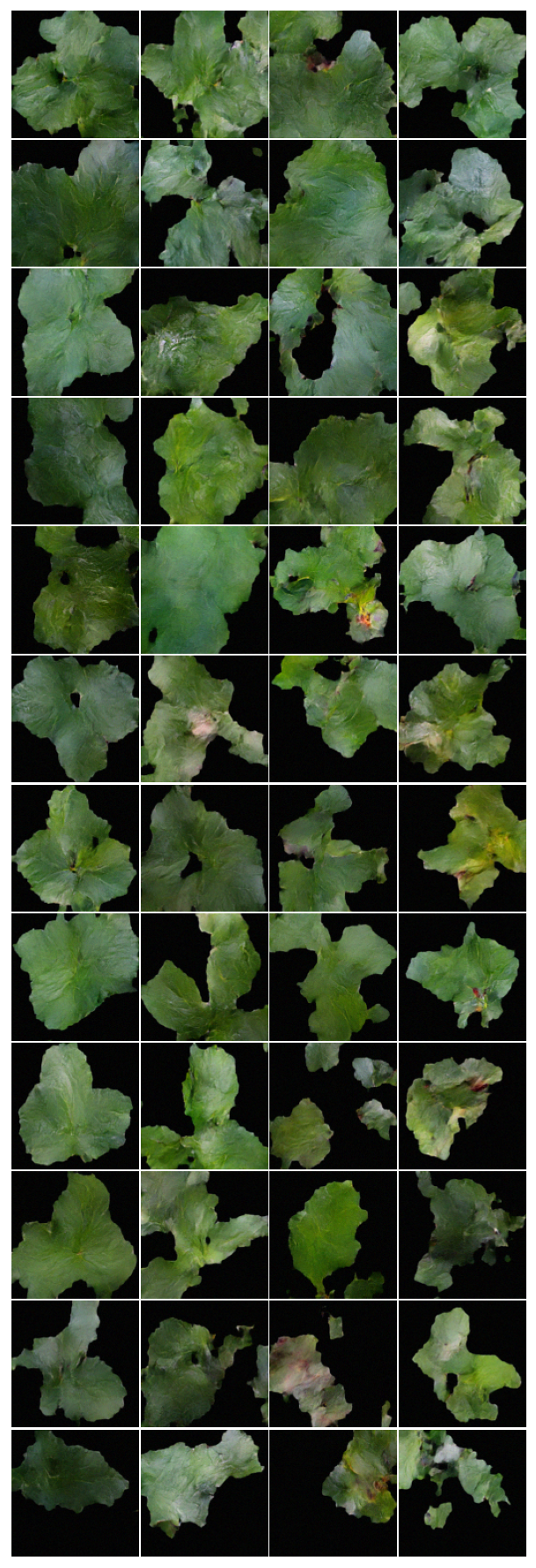}
		\caption{Generated images}
	\end{subfigure}
	\caption{Real images and generated images using the AP-CDE model for the leaves of strawberry plants. \label{fig:gene_straw}}
\end{figure}

\cref{tab:straw_den} shows the average bits per dimension results for all models on the training and testing sets. The proposed AP-CDE model outperforms the other competitors on density estimation, while gets results as good as Glow. \cref{fig:gene_straw}(b) shows generated images for each of the four classes. As a comparison, \cref{fig:gene_straw}(a) shows the real images for each of the four classes. The second column shows the generated powdery mildew images, which clearly have some white spots on the leaves; the third column shows the generated anthracnose images, which have purple spots on the leaves; and the fourth column shows the generated fusarium wilt images.

\begin{figure}[ht]
	\centering
	\captionsetup[subfigure]{justification=centering}
	\begin{subfigure}[t]{0.24\linewidth}
		\centering
		\includegraphics[width=1\linewidth]{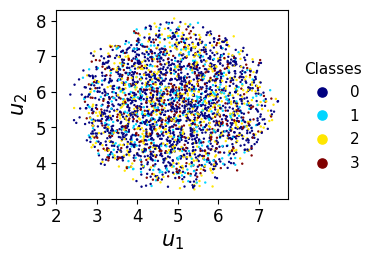}
		\caption{Glow}
	\end{subfigure}
	\begin{subfigure}[t]{0.24\linewidth}
		\centering
		\includegraphics[width=1\linewidth]{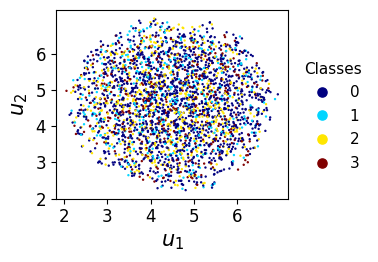}
		\caption{Glow-Mix}
	\end{subfigure}
	\centering
	\begin{subfigure}[t]{0.24\linewidth}
		\centering
		\includegraphics[width=1\linewidth]{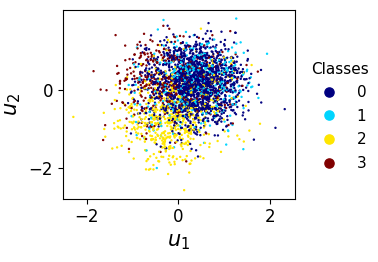}
		\caption{AP-CDE, the first two dimensions}
	\end{subfigure}
	\begin{subfigure}[t]{0.24\linewidth}
		\centering
		\includegraphics[width=1\linewidth]{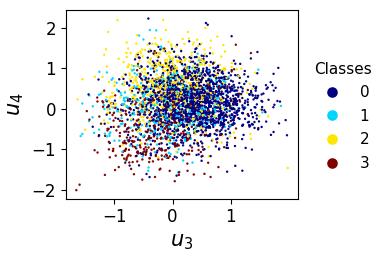}
		\caption{AP-CDE, the third and fourth dimensions}
	\end{subfigure}
	\caption[The latent variables mapping from the leaves of strawberry plants images produced from the models.]{The latent variables mapping from the leaves of strawberry plants images produced from the models. The UMAP is used to reduce the dimensions to 2 for Glow and Glow-Mix models. \label{fig:straw}}
\end{figure}

For this complex dataset, the proposed model still captures some features of the diseases of the strawberry plants, and gives $20.8\%$ classification error on the testing set, where the label is predicted via the logistic regression model on the $z_{P,i}$. We also validate the model using the method stated in \cref{subsec:dimreduction} and the independent classification model for which the ResNet101 is used gets $82.7\%$ accuracy on the new generated images. This means the AP-CDE model is helpful for conditional generating more images of leaves of strawberry plants.
\cref{fig:straw} plots the latent representations produced by these models. For Glow and Glow-Mix, the UMAP \citep{mcinnes2018umap} is used to reduce the dimension of latent $z_i$ and plot its output in 2D. For the latent $z_{P,i}$ produced by AP-CDE, the plot between the first and the second dimension is provided, as well as the plot between the third and the fourth dimension. The Glow and Glow-Mix models do not provide any useful dimension reduction, while the latent variables provided by AP-CDE are separated over the four classes.

\section{Discussion}

We extends the normalizing flow neural network to the task of CDE. It produces a generative model for high-dimensional data that can incorporate information from external predictors. Importantly, by using only a subset of the one-to-one transform from the high-dimensional data, a useful dimension reduction is achieved, in which the low-dimensional representation is empirically sufficient to characterize the changes of the response variable due to the predictor.

A number of neural network-based models for CDE have appeared in the literature (see, e.g., recent reviews \cite{ambrogioni2017kernel,rothfuss2019conditional}). Nevertheless, we want to emphasize the versatility and simplicity of the proposed approach. AP-CDE can work with any 
existing normalizing flow network architecture, with only a modification of the base distribution density from a normal one to the product of a prior distribution and a likelihood function.

There are several interesting directions for future work. First, there is a connection of the strategy for new photo synthesis---``keeping $z_P$, sampling $\tilde z_N$ and pulling back via $T^{-1}$''---to the popular practice of ``data augmentation in deep learning'' \citep{shorten2019survey}. Conventionally, to counter the small training sample problem, especially in image modeling, one relies on techniques such as geometric transformations, color space augmentation and random erasing. Nevertheless, there is a recent trend in using another neural network for data augmentation, such as adversarial training and neural style transfer. The proposed AP-CDE can be considered another  solution. Second, the normalizing flow networks can be too flexible, in the sense that they could transform a data distribution approximately to {\em any} latent distribution. This is likely  why the unsupervised mixture of Gaussian latent distribution fails to explain the variations in the observed space. The proposal of using the predictor-conditional distribution shows that there is room to make the latent variable more interpretable; nevertheless, caution should be taken and additional validation methods could be developed.

\bibliographystyle{plainnat}
\bibliography{reference.bib}

\appendix

\section{MNIST Handwritten Digit Images} \label{sec:mnist}

In this section, additional experiment results are shown using the MNIST data.

The MNIST data is used to illustrate the CDE when $x$ is discrete. This dataset contains $70,000$ processed images of handwritten digits, each having $28\times 28$ pixels. We use $60,000$ for training purposes and the remaining for testing. Each image $y_i$ is associated with a discrete label $x_i$ recording the ground-truth digit from $0$ to $9$.

The structure of this dataset is very similar to FashionMNIST \citep{xiao2017fashion} which is used in the main paper, and the same models as in Section 3.1 is used. The latent $z_{P,i}$ is chosen to be $z^{(3)}_{1:16,1,1}$. Unsurprisingly, the results corresponding to this dataset is quite similar to the ones from the FashionMNIST dataset.

\begin{figure}[ht]
	\centering
	\captionsetup[subfigure]{justification=centering}
	\begin{subfigure}[c]{0.32\textwidth}
		\centering
		\includegraphics[width=1\linewidth]{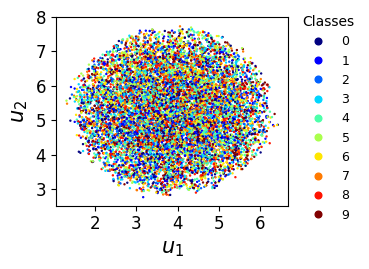}
		\caption{Glow}
	\end{subfigure}
	\begin{subfigure}[c]{0.32\textwidth}
		\centering
		\includegraphics[width=1\linewidth]{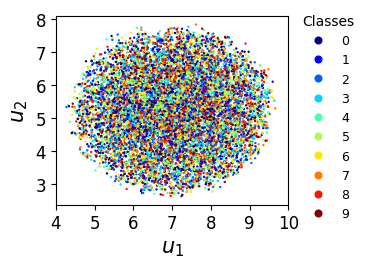}
		\caption{Glow-Mix}
	\end{subfigure}
	\begin{subfigure}[c]{0.32\textwidth}
		\centering
		\includegraphics[width=1\linewidth]{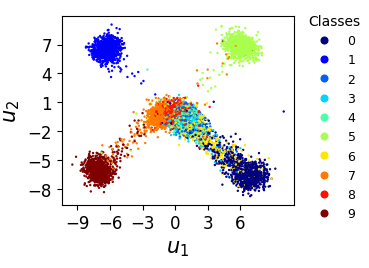}
		\caption{AP-CDE}
	\end{subfigure}
	\caption[Latent representations estimated by three models applied on the MNIST training set.]{Latent representations estimated by three models applied on the MNIST training set. For Glow and Glow-Mix models, since the latent are in high dimensions, the UMAP is used to reduce the dimensions to $2$. Neither of them produces interpretable latent presentations, whereas AP-CDE model does the latent that can be easily separated into groups. \label{fig:mnist}}
\end{figure}

Figure \ref{fig:mnist} plots the latent representations produced by Glow, Glow-Mix and AP-CDE. For AP-CDE, the latent $z^{(3)}_{1:2,1,1}$ is used as the $z_{P,i}$ for a better illustration in two dimensional reduction. For Glow and Glow-Mix, the UMAP \citep{mcinnes2018umap} is used to reduce the dimension and plot its output in 2D. For AP-CDE, the latent $z_{P,i}$ is provided. As expected, the latent variable produced by Glow follows a simple spherical Gaussian, and thus is not interpretable. The Glow-Mix does not produce a meaningful result either. Using AP-CDE and the supervising label information form $x_i$, we obtain a good separation of the ten digits based on the low-dimensional representation $z_{P,i}\in\mathbb{R}^2$ (shown in Figure \ref{fig:mnist}(c)). 

Under the working model using $z^{(3)}_{1:16,1,1}$ as $z_{P,i}$, for the testing set, we obtain $z_{P,i}$ and predict the label via the trained logistic regression model [using \(\arg\max_{k\in\{0,\ldots,9\}} f_{x\mid z_P}(k\mid T_{\hat\theta}^P(y_i))\)] to obtain classification accuracy of $98.5\%$. Empirically it is shown that the low-dimensional representation $z_{P,i}$ contains almost all the information to separate the different classes. As described in Section 3.1, we employ the ResNet101 to classify the artificially generated images from the proposed AP-CDE model with fixed $z_{P,i}$'s, and find that $96.14\%$ of them are still classified to the same class label as the $y_i$'s.  Hence, it is concluded that the $z_{N,i}$ largely corresponds to within-class variation, whereas $z_{P,i}$ captures between-class variation.

\begin{table}[ht!]
	\centering
	\caption{The averages of the bits per dimension on the training and the testing sets of MNIST data.} \label{tab:den0}
	\begin{tabular}{lrrrrr}\toprule
		Models    & Glow & Glow-Mix & Glow-CDE & Glow-NCDE & AP-CDE\\ \midrule
		Training set  & 1.01 & 1.02 &  1.03 & 1.23 & 1.01 \\
		Testing set   & 1.01 & 1.03 &  1.04 & 1.28 & 1.01 \\
		\bottomrule
	\end{tabular}
\end{table}

As seen in \cref{fig:mnist}(c), $x_i$ tends to make each group of $z_i$'s (each group corresponding to a digit) more concentrated to their respective center, which leads us to compare the marginal densities $f_y(y_i)$ with the ones from competing methods. \cref{tab:den0} depicts the average of the log-densities on the training and the testing sets for all models. Clearly, for this dataset, AP-CDE, likely due to a better group-wise concentration, produces overall higher or not lower marginal densities compared to its competitors. Note that bits per dimension is the negative log density over the number of dimensions.

\begin{figure}[ht]
	\centering
	\captionsetup[subfigure]{justification=centering}
	\begin{subfigure}[t]{0.35\textwidth}
		\centering
		\includegraphics[width=1\linewidth]{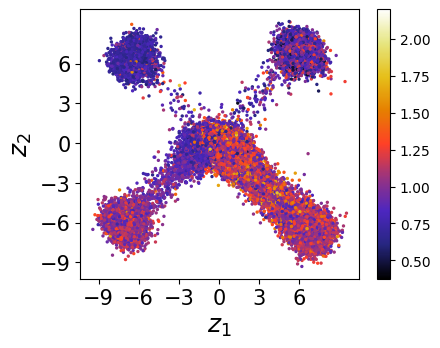}
		\caption{Training set.}
	\end{subfigure}
	\begin{subfigure}[t]{0.35\textwidth}
		\centering
		\includegraphics[width=1\linewidth]{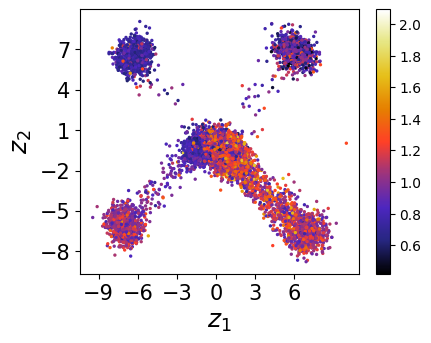}
		\caption{Testing set.}
	\end{subfigure}
	\caption{The first two dimensions of the latent variables from AP-CDE, colored by the estimated densities in the scale of bits per dimension. \label{fig:mnist_den}}
\end{figure}

\begin{figure}[ht]
	\centering
	\includegraphics[width=0.65\linewidth]{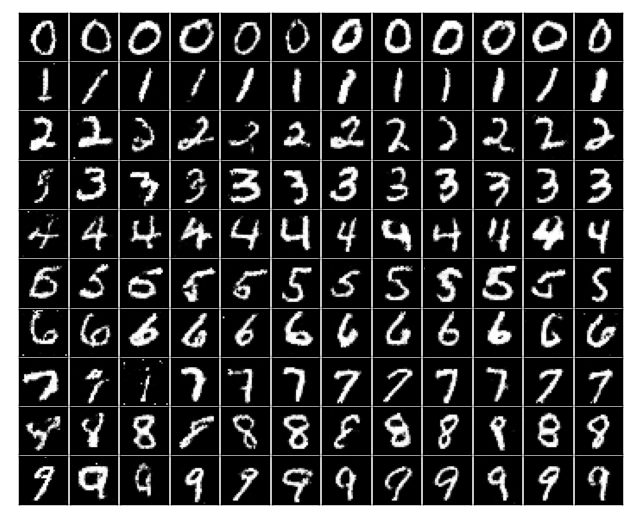}
	\caption{Sample images from the MNIST data, with each row sorted in the increasing order of estimated densities. \label{fig:gene_mnist}}
\end{figure}

Further, the latent $z_{P,i}$ is colored using the magnitude of the bits per dimension (\cref{fig:mnist_den}). In the result, those points with relatively high BPM values (low density values) tend to correspond to images of low quality or higher ambiguity regarding the digit class. The easier classified classes have relatively higher densities. To show this, we plot a few sampled digits in \cref{fig:gene_mnist} and sort each row by the density value in increasing order. It can be seen that the images on the left tend to be harder to assign to a class, compared to the ones on the right.

\end{document}